\documentclass[aps,pra,superscriptaddress,floatfix,showpacs,showkeys,10pt,twocolumn]{revtex4}
\usepackage{amssymb}
\usepackage{amsfonts}
\usepackage{amsmath}
\usepackage[dvips]{graphicx}
\usepackage{amsmath,amsfonts,amssymb,graphics,graphicx,epsfig,color,times,bbm}

\begin{document}

\title{Discrete quantum Fourier transform in coupled semiconductor double quantum dot molecules}
\author{Ping Dong \footnote{Email: dongping9979@163.com}}
\author{Ming Yang}
\affiliation{Key Laboratory of Opto-electronic Information
Acquisition and Manipulation, Ministry of Education, School of
Physics {\&} Material Science, Anhui University, Hefei, 230039, P R
China}
\author{Zhuo-Liang Cao \footnote{Email: zhuoliangcao@gmail.com,
Telephone: 086-551-5108049}}

\affiliation{Key Laboratory of Opto-electronic Information
Acquisition and Manipulation, Ministry of Education, School of
Physics {\&} Material Science, Anhui University, Hefei, 230039, P R
China} \affiliation{Department of Physics, Hefei teacher college,
230061, P R China}

\begin{abstract}
In this letter, we present a physical scheme for implementing the
discrete quantum Fourier transform in a coupled semiconductor double
quantum dot system. The main controlled-R gate operation can be
decomposed into many simple and feasible unitary transformations.
The current scheme would be a useful step towards the realization of
complex quantum algorithms in the quantum dot system.
\end{abstract}


\maketitle

The solid-state quantum computation attracts many interests since
solid systems are more easily integrated into large quantum
networks. Semiconductor quantum dots are considered to be one of the
most promising candidates for quantum computation in solid state
\cite{1}. Recently, qubits, encoded on the electron-spin singlet
state
$|S\rangle=(|\uparrow\downarrow\rangle-|\downarrow\uparrow\rangle)/\sqrt{2}$
and triplet state
$|T\rangle=(|\uparrow\downarrow\rangle+|\downarrow\uparrow\rangle)/\sqrt{2}$
of coupled double quantum dot (DQD) molecules, which are being
widely researched in theory and experiment
\cite{2,3,4,5,6,7,8,9,10,11,12,13}. It is shown that the coherent
time of electron-spin state is comparatively long comparing with the
state of charges, and the qubits can be protected from low-frequency
noise and can suppress the dominant source of decoherent from
hyperfine interaction. Due to their advantages, many schemes based
on the two spin states in coupled DQD molecules have been proposed,
such as the realization of Bell-state measurement \cite{14} and the
generation of cluster states \cite{15}, etc.

\begin{figure}[tbp]
\includegraphics[scale=0.76,angle=0]{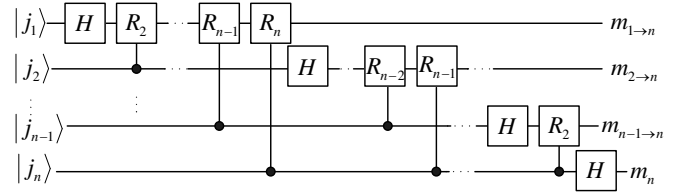}
\caption{The effective circuit diagram of discrete QFT \cite{18}.
$j_{l}$ ( $l=1,2,\cdots,n$) are inputs and $m_{l \rightarrow
n}=(|0\rangle+e^{2\pi i0.j_{l}\cdots j_{n}}|1\rangle)/\sqrt{2}$ are
outputs. $R_{\kappa}$ ($\kappa=2,3,\cdots,n$) are a series of phase
transformations and $H$ is the Hadamard transformation. The black
dots present the control bits.}\label{fig1}
\end{figure}

Inspired by above ideas, we propose a scenario for the
implementation of discrete quantum Fourier transform (QFT) via
coupled QDQ molecules arranged in line. As we known, only a few
physical schemes of QFT have been proposed, one based on cavity
quantum electrodynamics (QED) \cite{16} and another based on nuclear
magnetic resonance (NMR) \cite{17}. QFT does not speed up the
classical tasks of classical Fourier transform (CFT), however, it
provides a first step towards the implementation of Shor's factoring
and other quantum algorithms, \emph{i.e.}, it is the key ingredient
for order-finding problem, factoring problem, counting solution, the
solution of hidden subgroup problem and so on. Therefore QFT is
still very important in quantum computation. For discrete CFT of $N$
inputs $x_{j}$ ($j=0,1,2,\cdots,N-1$), the outputs $y_{k}$
($k=0,1,2,\cdots,N-1$) can be expressed as
$$y_{k}\equiv \frac{1}{\sqrt{N}}\sum^{N-1}_{j=0}x_{j}e^{2\pi
ijk/N}.$$ Correspondingly, the discrete QFT is defined as
$$|j\rangle\xrightarrow[]{QFT}\frac{1}{\sqrt{N}}\sum^{N-1}_{k=0}e^{2\pi
ijk/N}|k\rangle,$$ where $|k\rangle$ is a set of normative
orthogonal basis.  For simplicity, the transformation can be
rewritten as the following form
\begin{eqnarray}
&|j_{1}&|j_{2}\cdots |j_{n}\rangle  \xrightarrow[] {U_{QFT}} \nonumber\\
&&(|0\rangle+e^{2\pi i0.j_{n}}|1\rangle)(|0\rangle+e^{2\pi
i0.j_{n-1}j_{n}}|1\rangle)\cdots \nonumber\\&&(|0\rangle+e^{2\pi
i0.j_{2}\cdots j_{n}}|1\rangle)(|0\rangle+e^{2\pi
i0.j_{1}j_{2}\cdots j_{n}}|1\rangle)/2^{n/2} \nonumber.
\end{eqnarray}
where $0.j_{1}j_{2}\cdots j_{n}=j_{1}/2+j_{2}/4+ \cdots +
j_{n}/2^{n}$, $j_{l}$ ( $l=1,2,\cdots,n$) equal to 0 or 1. The
effective circuit diagram of discrete QFT is shown in Fig. \ref
{fig1}. The QFT is a unitary transformation, which is $
 R_{\kappa}=\left(
        \begin{array}{cc}
         1 & 0 \\
          0 & e^{2\pi i/2^{\kappa}}
           \end{array}
          \right).$

\begin{figure}[tbp]
\includegraphics[scale=0.92,angle=0]{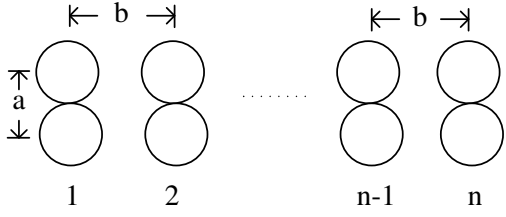}
\caption{Schematic diagram of coupled DQD molecules (the $n$
identical molecules arranged in line). The hollow circle presents a
quantum dot with one electron and two hollow circles construct a
molecule. The distance between two dots of each molecule is $a$ and
the distance between two nearest-neighbor molecules is $b$.
}\label{fig2}
\end{figure}

Next, we investigate a detailed scenario for implementing the
discrete QFT via coupled DQD molecules. There are $n$ semiconductor
coupled DQD molecules (GaAs) arranged in line, which is shown in Fig
\ref {fig2}. Each molecule includes two electrons. According to the
explicit analysis of Refs. \cite{14,15,19}, the charge states of
each molecule (0,2), (1,1) and (2,0) can be transferred by sweeping
the bias parameter $\Delta$. For ($n_{1},n_{2}$), $n_{1}$ ($n_{2}$)
denotes the number of electrons in the upper (lower) quantum dots,
and (0,2), (1,1) and (2,0) respectively correspond to $\Delta=E$, 0
and $-E$. $E$ is the charging energy of each quantum dot. $\Delta$
can be controlled by gate-bias voltages of each molecule or by
external electrical field. It is noted that the triplet state
$|T\rangle$ will be in the charge state (1,1) and the singlet state
$|S\rangle$ could be in (0,2), (1,1) or (2,0) if the initial charge
state is in (1,1) during the process of sweeping bias parameter
because of Pauli blockade.

In order to implement the discrete QFT, we prepare the $n$ DQD
molecules in spin states $|S\rangle=|0\rangle$ or
$|T\rangle=|1\rangle$ randomly, and adjust $\Delta=0$ (the charge
states are all in (1,1)). Thus the total spin state is
$|\psi\rangle=|j_{1}j_{2}\cdots j_{n}\rangle$, $j_{l}=|S\rangle$ or
$|T\rangle$ ($l=1,2,\cdots, n$). The neighbor two DQD molecules are
separated by an impenetrable barrier, so we only need consider the
Coulomb interaction between two molecules. The detailed process can
be described by the following $n$ steps:

\emph{\textbf{The first step}}: Firstly, we perform single-qubit
gate transformations on DQD molecules 1 and 2, respectively,
\begin{equation}
\label{1}
|0\rangle_{1}\rightarrow\frac{1}{\sqrt{2}}(|0\rangle_{1}+|1\rangle_{1}),
|1\rangle_{1}\rightarrow\frac{1}{\sqrt{2}}(|0\rangle_{1}-|1\rangle_{1}),
\end{equation}
and
\begin{equation}
\label{2} |0\rangle_{2}\rightarrow|0\rangle_{2},
|1\rangle_{2}\rightarrow
e^{i\theta/2}|1\rangle_{2},(\theta=\theta_{12}=\pi/2)
\end{equation}
and a single-qubit operation on DQD molecules 1 as in  Eq.
(\ref{1}), which can be achieved by Euler angle method or by $U_{Z}$
rotations and $U_{XZ}$ rotations in the XZ plane with finite
singlet-triplet energy splitting \cite{12}. Then we sweep the bias
parameter $\Delta$ of molecules 1 and 2, the effect Hamiltonian is a
Ising model, which can be expressed as \cite{15,19}
\begin{equation}
\label{3}
H=E_{12}\frac{1-\sigma^{1}_{z}}{2}\frac{1-\sigma^{2}_{z}}{2},
\end{equation}
where
$E_{12}=\frac{1}{4\pi\epsilon}(\frac{2e^{2}}{b}-\frac{2e^{2}}{\sqrt{a^{2}+b^{2}}})$,
$\epsilon$ is the dielectric constant of GaAs and $\sigma_{z}$ is a
Pauli operator. In this process, we adjust the interaction time
$t_{12}$, let $E_{12}t_{12}/\hbar=(2\mu+1)\pi$, ($\mu\in \mathcal
{\mathbb{N}}$), thus we have the evolution
\begin{eqnarray}
\label{4}
|00\rangle_{12}\rightarrow|00\rangle_{12},
|01\rangle_{12}\rightarrow|01\rangle_{12},\nonumber\\
|10\rangle_{12}\rightarrow|10\rangle_{12},
|11\rangle_{12}\rightarrow -|11\rangle_{12}.
\end{eqnarray}
Then we perform the single-qubit gate transform of Eq. (\ref{1}),
and another single-qubit transform
\begin{equation}
\label{5} |0\rangle_{1}\rightarrow|0\rangle_{1},
|1\rangle_{1}\rightarrow e^{-i\theta/2}|1\rangle_{1},
(\theta=\theta_{12}=\pi/2)
\end{equation}
and the single-qubit transform of Eq. (\ref{1}) again. We again weep
$\Delta$ of molecules 1 and 2 to drive the interaction as in Eq.
(\ref{3}-\ref{4}), perform the single-qubit transform of  Eq.
(\ref{1}) and single-qubit transform as in Eq. (\ref{2}) on molecule
1.

Secondly, we mainly consider the operations on molecules 1 and 3,
the process is similar as above mentioned (from Eq. (\ref{2}) to the
end of above paragraph), in which
$\theta=\theta_{13}=\pi/2^{2}=\pi/4$,
$E_{13}=\frac{1}{4\pi\epsilon}[\frac{2e^{2}}{2b}-\frac{2e^{2}}{\sqrt{a^{2}+(2b)^{2}}}]$.
Similarly, we consider the molecules 1 and 4, molecules 1 and 5, up
to molecules 1 and $n$ one by one. In a word,  the angle  $\theta$
satisfies $\theta=\theta_{1\iota}=\pi/2^{\iota-1}$,  the energy
$E_{1\iota}$ will be
$E_{1\iota}=\frac{1}{4\pi\epsilon}[\frac{2e^{2}}{(\iota-1)b}-\frac{2e^{2}}{\sqrt{a^{2}+(\iota-1)^{2}b^{2}}}]$,
and the interaction time satisfies
$E_{1\iota}t_{1\iota}/\hbar=(2\mu+1)\pi$, ($\iota=2,3,\cdots,n$).
The state of molecule 1 becomes $m_{1\rightarrow
n}=(|0\rangle+e^{2\pi i0.j_{l}\cdots j_{n}}|1\rangle)/\sqrt{2}$.

\emph{\textbf{The second step}}: Firstly, we still perform the
single-qubit transformation on molecule 2 as in Eq. (\ref{1}). Then
considering the operations on molecules 2 and 3, 2 and 4, $\cdots$,
2 and $n$ as the process of the first step. Here,
$\theta=\theta_{2\iota}=\pi/2^{\iota-2}$,
$E_{2\iota}=\frac{1}{4\pi\epsilon}[\frac{2e^{2}}{(\iota-2)b}-\frac{2e^{2}}{\sqrt{a^{2}+(\iota-2)^{2}b^{2}}}]$,
and $E_{2\iota}t_{1\iota}/\hbar=(2\mu+1)\pi$,
($\iota=3,4,\cdots,n$). The state of molecule 2 becomes
$m_{2\rightarrow n}=(|0\rangle+e^{2\pi i0.j_{2}\cdots
j_{n}}|1\rangle)/\sqrt{2}$.

\textbf{$\cdots\cdots\cdots$}

\emph{\textbf{The last step}}: We only need perform a single-qubit
transformation on molecule $n$ as in Eq. (\ref{1}). The state of
molecule $n$ becomes $m_{n}=(|0\rangle+e^{2\pi
i0.j_{n}}|1\rangle)/\sqrt{2}$.

Finally, we read out the total result in reversed order, i.e., from
molecule $n$ to molecule 1. The result can be expressed as
\begin{eqnarray}
|\Phi\rangle_{total} &=&(|0\rangle+e^{2\pi
i0.j_{n}}|1\rangle)_{n}(|0\rangle+e^{2\pi
i0.j_{n-1}j_{n}}|1\rangle)_{n-1}\nonumber\\&& \cdots
(|0\rangle+e^{2\pi i0.j_{1}j_{2}\cdots j_{n}}|1\rangle)_{1}/2^{n/2},
\end{eqnarray}
which is the result of standard discrete QFT. By far, we have
completed the task successfully. It is shown that, during the whole
process, $\theta=\theta_{\gamma \iota}=\pi/2^{\iota-\gamma}$ and
$E_{\gamma\iota}=\frac{1}{4\pi\epsilon}[\frac{2e^{2}}{(\iota-\gamma)b}-\frac{2e^{2}}
{\sqrt{a^{2}+(\iota-\gamma)^{2}b^{2}}}]$, where $\gamma=1,2,\cdots,
n-1$.

Initialization of the system is an important task for quantum
computation. In the current scheme, the arbitrary initial state
($|S\rangle$ or $|T\rangle$) needs to be generated for an arbitrary
QFT, however, for the practical QFT, the initial state must be a
known one. In addition, the states $|S\rangle$ and $|T\rangle$ can
be transformed each other. So we only need initialize the system to
the state $|S\rangle$. As described in Ref. \cite{14,15}, it can be
realized by loading two electrons from a nearby Fermi sea into the
ground state of a single quantum dot and then sweeping the bias
parameter $\Delta$ from $E$ to $-E$ by the rapid adiabatic passage
to change state from (0,2) to (1,1). For the one-dimensional qubit
array, we need at least two steps to initialize all qubit to the
state $|S\rangle$  because the initialization can not be made on
non-neighboring qubits simultaneously \cite{14}.

Next, we discuss the feasibility  of the current scheme. Generally,
the coherent time of electronic spin state will be affected by the
hyperfine interaction between spins and nucleus. We choose the
singlet and triplet spin states as qubits, which can weaken the
negative effect efficiently. Charge fluctuations in environment, an
important resource of decoherent for charge qubits in semiconductor
\cite{a}, which can lead to gate errors and dephase \cite{3}. The
unavoidable background charge noise, nuclear-spin-related noise and
control electrical noise will affect the tunnel coupling between
double quantum dots and the effective interaction strength between
different molecules \cite{15,b}. All of above noise will result in
an unwanted phase $\delta\phi$. Assume that the $\delta\phi$ has a
Gaussian distribution $G(0,\sigma)$ with average value of zero and
variance of $\sigma$ \cite{15,c}. According to Ref. \cite{15,c}, we
can see that the fidelity of the Controlled-phase gate will be above
$96\%$ in this interaction model with $\delta\phi\simeq \pm
0.03\pi$. Moreover, this kind of noise can also be mended by
adjusting the gate voltage of every molecule \cite{d}.

The spin coherent time can reach $1.2 \mu s$ by using  spin-echo
technology \cite{13}. In our scheme, the process mainly includes the
controlled-R gate transformation. If the smallest Coulomb energy
reduces to 0.1 percent of the largest Coulomb energy, \emph{i.e.},
$\frac{E_{min}}{E_{max}}=\frac{\frac{1}{4\pi\varepsilon}(\frac{2e^{2}}{nb}-\frac{2e^{2}}{\sqrt{a^{2}+(nb)^{2}}})}
{\frac{1}{4\pi\varepsilon}(\frac{2e^{2}}{b}-\frac{2e^{2}}{\sqrt{a^{2}+b^{2}}})}=0.1\%$,
the number of the DQD molecule will be $n\simeq16$. Assume that $a=5
nm$ and $b=12nm$, the total interaction time of the whole process
will be $t_{total}\simeq 20 ns$. The ratio is only $1.6\%$ between
the total interaction time and coherent time. Therefore we can
complete the discrete QFT in the coherent time scale.

We should also pay attention to the process of interactions between
two molecules. In order to avoid interaction noise coming from
others quantum dot molecules, we can consider the electronic shield
technology, \emph{i.e.}, if one needs quantum dot molecules A and B
to interact each other, others can be all shielded by electric
shielding boxes put to earth. In this case, arbitrary two molecules
interact each other with a different distance, and the interaction
of them is governed by Eq. (\ref{3}). A difficulty in experiment
would be that the electric shielding box will slightly affect the
interaction of the two DQD molecules with a certain extent.  In
addition, we separate any two neighbor molecules by impenetrable
barriers, which can prevent the tunneling effect from two different
molecules each other.

In conclusion, we present a scheme for implementing the discrete QFT
via coupled semiconductor DQD molecules (GaAs). The main task is to
realize controlled-R transformations $U(R)$. We can rewrite it by
product form with some unitary transformations, i.e.,
$U_{12}(R)=U_{2}(\frac{\theta}{2})U_{cnot}U_{2}(-\frac{\theta}{2})U_{cnot}U_{1}(\frac{\theta}{2})$
or
$U_{12}(R)=U_{2}(\frac{\theta}{2})U_{2}(H)U_{c-z}U_{2}(H)U_{2}(-\frac{\theta}{2})
U_{2}(H)U_{c-z}U_{2}(H)U_{1}(\frac{\theta}{2})$. Every unitary
transformation can be simply realized in our scheme. It is very
important for the realization of universal quantum computation.
Meanwhile, discrete QFT is a key ingredient for some quantum
algorithms. Therefore, our scheme would be a useful step towards the
realization of complex quantum algorithms in quantum dot system.

\begin{acknowledgments}
This work is supported by National Natural Science Foundation of
China (NSFC) under Grant Nos: 60678022 and 10704001, the Specialized
Research Fund for the Doctoral Program of Higher Education under
Grant No. 20060357008, Anhui Provincial Natural Science Foundation
under Grant No. 070412060, the Talent Foundation of Anhui
University, Anhui Key Laboratory of Information Materials and
Devices (Anhui University).

\end{acknowledgments}


\begin{thebibliography}{99}
\bibitem{1} D. Loss, and D. P. DiVincenzo, Phys. Rev. A 57, 120
(1998).
\bibitem{2} X. D. Hu, and S. DasSarma, Phys. Rev. A 61, 062301
(2000).
\bibitem{3} G. Burkard, D. Loss, and D. P. DiVincenzo, Phys. Rev. B
59, 2070 (1999).
\bibitem{4} J. Schliemann, D. Loss, and A. H. MacDonald, Phys. Rev. B
63, 085311 (2001).
\bibitem{5} I. A. Merkulov, Al. L. Efros, and M. Rosen, Phys. Rev. B
65, 205309 (2002).
\bibitem{6} R. Kotlyar, C. A. Stafford, and S. DasSarma, Phys. Rev.
B 58, 3989 (1998).
\bibitem{7} J. R. Petta, A. C. Johnson, J. M. Taylor, E. A. Laird, A.
Yacoby, M. D. Lukin, C. M. Marcus, M. P. Hanson, and A. C. Gossard,
Science 309, 2185 (2005).
\bibitem{8} J. M. Taylor, H. A. Engel, W. D\"{u}r, A. Yacoby, C. M.
Marcus, P. Zoller, and M. D. Lukin, Nature Physics 1, 177 (2005).
\bibitem{9} A. C. Johnson, J. R. Petta, J. M. Taylor, A.
Yacoby,  M. D. Lukin, C. M. Marcus, M. P. Hanson, and A. C. Gossard,
Nature 435, 935 (2005).
\bibitem{10} J. M. Taylor, W. D\"{u}r, P. Zoller, A. Yacoby, C. M.
Marcus, and M. D. Lukin, Phys. Rev. Lett. 94, 236803 (2005).
\bibitem{11} D. Stepanenko, and G. Burkard, Phys. Rev. B 75, 085324 (2007).
\bibitem{12} R. Hanson, and G. Burkard, Phys. Rev. Lett. 98, 050502
(2007).
\bibitem{13} J. M. Taylor, J. R. Petta, A. C. Johnson, A. Yacoby,  C. M.
Marcus, and M. D. Lukin, \emph{cond-mat/0602470}.
\bibitem{14} H. Zhang, G. P. Guo, T. Tu, and G. C. Guo, Phys. Rev. A 76,
012335 (2007).
\bibitem{15} G. P. Guo, H. Zhang, T. Tu, and G. C. Guo,
 Phys. Rev. A 75, 050301(R) (2007).
\bibitem{16} M. O. Scully, and M. S. Zubairy, Phys. Rev. A 65, 052324
(2002).
\bibitem{17} Y. S. Weinstein, M. A. Pravia, E. M. Fortunato, S. Lloyd, and
D. G. Cory, Phys. Rev. Lett. 86, 1889 (2001).
\bibitem{18} M. A. Nielsen, and I. L. Chuang, \emph{Quantum Computation
and Quantum Information}, Cambridge University Press, chapter 5,
2000.
\bibitem{19} G. P. Guo, X. J. Hao, T. Tu, Z. C. Zhu, and G. C. Guo, Eur. Phys. J. B 61, 141 (2008).
\bibitem{a} T. Hayashi,  T. Fujisawa, H. D. Cheong, Y. H. Jeong, and Y.
Hirayama, Phys. Rev. Lett. 91, 226804 (2003).
\bibitem{b} X. D. Hu, and S. Das Sarma, Phys. Rev. Lett. 96, 100501 (2006).
\bibitem{c} M. S. Tame, M. Paternostro, M. S. Kim, and V. Vedral,
Int. J. Quantum Inf. 4, 689 (2006).
\bibitem{d}  T. Tanamoto, Y. X. Liu, S. Fujita, X. D. Hu, and F.
Nori, Phys. Rev. Lett. 97, 230501 (2006).
\end{thebibliography}
\end{document}